\documentstyle[12pt]{article}
\def \partialslash{\!\!\not{\!\partial}}
\def \Aslash{\!\!\not{\!\! A}}
\def \kslash{\!\!\not{\! k}}
\def \qslash{\!\!\not{\! q}}

\begin{document}

preprint Coimbra 940506

IJS-TP-94/10

\begin{center}

{\large \bf MOMENTUM DEPENDENT VERTICES
$\sigma \gamma \gamma$, $\sigma \rho \gamma$
and $\sigma \rho \rho$ : THE NJL SCALAR HIDDEN BY CHIRAL SYMMETRY}
\footnote{Work supported in part by CERN no. PCERN/C/FAE/74 /91,
CNPq, STRDA/C/CEN/564/92
and the Ministry of Science
and Technology of the Republic of Slovenia.}

\vspace{2.5cm}

B. Bajc$^{\dag}$, A. H. Blin$^{\S}$, B. Hiller$^{\S}$,
M.C. Nemes$^*$ and M. Rosina$^{\dag \ddag}$

\vspace{0.5cm}

$^{\dag}$Institut Jo\v zef Stefan, Jamova 39, p.p.100,
61111 Ljubljana, Slovenia

\vspace{0.5cm}

$^{\S}$Centro de F\'{\i}sica Te\'{o}rica, Dpt. de F\'{\i}sica,
P-3000 Coimbra, Portugal

\vspace{0.5cm}

$^*$Dpt. de F\'{i}sica, ICEX, Universidade Federal de Minas Gerais,
C.P.702, 31270 Belo Horizonte-MG, Brasil

\vspace{0.5cm}

$^{\ddag}$Dpt. of Physics, University of Ljubljana,
Jadranska 19, p.p.64, 61111 Ljubljana, Slovenia

\end{center}

\vspace{3cm}

ABSTRACT: We calculate the momentum dependence of
three particle vertices $\sigma \gamma \gamma$,
$\sigma \rho \gamma$ and $\sigma \rho \rho$
in the context of a Nambu Jona Lasinio type model.
We show how they influence the processes
$\gamma \gamma \rightarrow \sigma \rightarrow \pi \pi$,
$\rho \rightarrow \gamma \sigma$ and $\gamma \gamma
\rightarrow \rho \rho$ and how chiral symmetry
shadows
the presence of the $\sigma$.
\vfill

\pagebreak

\section{INTRODUCTION}

\vspace{0.5cm}

Presently most of the properties of low energy hadron
physics are described in terms of effective lagrangians
which incorporate some of the important symmetries of QCD.
The Nambu Jona Lasinio (NJL) type lagrangians are
particularly popular due to the fact that they provide
for an adequate description of low energy meson masses,
their decays, scattering, etc. in terms of quarks
\cite{1}-\cite{2b}. One of the important
predictions of the NJL model is the existence of a
broad scalar excitation - the $\sigma$ particle.
Its mass is exactly equal to $2 M_{u}$ ($M_{u}$ is
the constituent up-quark mass) in the chiral
limit (current quark masses $m_{u}=m_{d}=0$
or, equivalently, $m_{\pi}=0$) of the 2 flavours NJL
and it is a little bit smaller than $2 M_{u}$ in a
more general framework ($m_{\pi}\ne 0$, strange
quark contribution included and $U(1)_A$ symmetry
explicitly broken), which we will adopt. The
experimental evidence of the $\sigma$ resonance is
however controversial. Since its non-existence would enter in contradiction
with the NJL
model, it is one of the purposes of the present article to show
that the experimental results on the $\gamma \gamma
\to \pi \pi$ and $\rho \to \gamma \pi \pi$
processes are by no means in conflict with the presence
of the $\sigma$ meson predicted by the NJL. Moreover, the $\sigma$ meson
may resolve some puzzles connected with the
reactions $\gamma \gamma \to \rho \rho \to 4 \pi$ and
$ p p \to p p \pi^{0}$, as discussed in the sequel.

The amplitude analysis \cite{morgan} of the data
\cite{exppi+pi-}-\cite{exppi0pi0} for the reaction
$\gamma \gamma \to \pi \pi$ shows that the $\sigma$-
contribution in the s-channel near the pole cannot
be large. The same can be said for the decay
$\rho^0 \to \gamma \sigma$ \cite{dolinsky}: the
branching ratio for $\rho^0 \to \gamma \pi^+ \pi^-$
is in fact 1 \%, but the absence of $\rho^0 \to
\gamma \pi^0 \pi^0$ clearly means that the two pions
cannot be in a S-wave. It will be shown that the intermediate $\sigma$
meson in reactions involving photons is hidden due to the  chiral
symmetry prediction $m_{\sigma}=2M_u$.

The reaction
$\gamma \gamma \rightarrow \rho \rho \rightarrow 4\pi$
shows a large enhancement of the
$\rho ^0\rho ^0 $ \cite{3a} intermediate
channel as compared to the $\rho ^+\rho ^- $
\cite{3b} channel. Several
models give partial answers to the problem.
Within the MIT bag quark model,
the candidates for this reaction
appear as 2$q - 2\bar{q}$ states with masses
around 1.65 GeV \cite{4}. Using the multiplets
suggested in \cite{4}, but with a little lower
masses ($1.4$ GeV), the $\rho ^0\rho ^0 $ channel
is quite well reproduced, while the
so obtained $J^{P}=2^{+}$
above $1.7$ GeV and all the $0^{+}$ contributions
to the $\rho^+\rho^-$ channel are
too small \cite{5}. In the effective
lagrangian approach \cite{6}, using $\pi $, $\eta$,
$\eta'$, $\rho $, $\omega$, $\phi$, $\epsilon(1300) $,
$f_2(1270) $, and $a_2(1320)$, the K-matrix
method is employed to unitarize the amplitude. Similar
to the quark-inspired model \cite{5}, the
$\rho^{0} \rho^{0}$ cross section is well
reproduced, while the cross section obtained
for  $\rho^+\rho^- $
is too large by a factor of two.

We propose another
line of investigation, namely the hypothesis that the
reaction proceeds via a scalar exchange, because
the two photons would be converted only in
$\rho^{0}\rho^{0}$, due to zero charge exchange.
We are encouraged to
consider this mechanism by a simple qualitative
estimate in which we adapt the Yukawa-type scalar
exchange potential proposed by Machleidt et al.\cite{7}
to the $\rho \rho  $ system by multiplying with a
factor $(\frac{2}{3})^2$ since there are two quarks
rather than three at each vertex. Choosing a
reasonable hard core at $.17$ fm and solving
a non-relativistic Schr\" odinger equation, we obtain
at $E \approx 0$ a weakly bound or resonant state.
Our calculation
of the $q^2$ dependence of the form factors for
$\sigma \rho \gamma$ and $\sigma \rho \rho$
shows that the existence of the $\sigma$-meson
could resolve the puzzle offered. In this case
the $\sigma \rho \gamma$ vertex contributes in
a kinematical region far from the one where chiral
symmetry predicts shadowing of the scalar.

To complete the picture of the scalar meson predicted
by the NJL model, we mention its relevance also in
cases which do not involve photons and where
no shadowing due to chiral
symmetry is observed. One example is $\pi \pi$
scattering in some models
\cite{maria}, where the inclusion of the scalar is crucial for
the description of the data on scattering lengths and phase shifts. The
other example is the reaction $ p p \to p p \pi^{0}$ where the
hypothetical $\sigma$ meson can also
resolve a puzzle, namely the underestimate of the cross
section \cite{meyer}. As pointed out by Horowitz et al.
\cite{horowitz}, the inclusion of the $\sigma$ exchange
between a positive and a negative energy virtual state of
the intermediate protons can in fact account for the
discrepancy by a factor of $5$ between experimental
results and  theoretical expectations.

\vspace{0.5cm}
\section{THE MODEL AND FORM FACTORS}

\vspace{0.5cm}

The following NJL lagrangian has been shown to
provide for an adequate description of mesonic
mass spectra and radiative decays [5,18]

\begin{eqnarray}
{\cal L}=& &\bar{\psi}[\, i \partialslash
-m_{curr}
          -\frac{ie}{2} (\lambda _3 +
\frac{\lambda _8}{\sqrt{3}} )
\, \Aslash ]\psi \nonumber \\
 &+&G_{1}[(\bar{\psi} \lambda_i \psi)^{2}
+(\bar{\psi} i \gamma_{5} \lambda_{i} \psi)^{2}] \nonumber \\
 &+&G_2[(\bar{\psi} \lambda_{a}^c \gamma_{\mu} \psi)^{2}
+(\bar{\psi} \lambda_{a}^c \gamma_{\mu} \gamma_{5} \psi)^{2}]
\nonumber \\
 &+&K[{\rm det}\{\bar{\psi} (1+\gamma_{5}) \psi\}+{\rm det}\{\bar{\psi}
(1-\gamma_{5}) \psi\}]
\label{e1}
\end{eqnarray}

\noindent
which is an SU(3) generalization \cite{2} of the NJL model \cite{1},
coupled minimally to the electromagnetic interaction.
\noindent
Here the flavor index i runs from 0 to 8 with
$\lambda _0 = \sqrt{\frac{2}{3}} {\bf 1}$.
The $\lambda_a^c$ are color matrices ($a$=1,...,8) and
$m_{curr}$ is the diagonal current quark mass matrix.
The usual regularization scheme is used
with a covariant four-momentum cutoff,
$\Lambda =1$ GeV. The parameters used are the same as in
\cite{2b}: $G_{1}\Lambda^{2}=3.95$, $G_{2}\Lambda^{2}=5.43$,
$K \Lambda^{5}=42$ and the quark current masses
$m_{u}=m_{d}=4$ MeV, $m_{s}=115$ MeV.
The constituent quark masses $M_{i}$ for the given
model parameters are $M_{u}=M_{d}=390$MeV
and $M_{s}=536$MeV.
In the present calculation we are interested
in vertices of the type $\sigma VV'$, where $V$, $V'$ stand
for vector mesons or photons and $\sigma $ is the scalar
composed of $\lambda_0 - \lambda_8 $ flavor admixtures.
These vertices are obtained by evaluating the Feynman
diagram shown in fig.1 , where the triangle is composed
of quark lines. The full electromagnetic vertex, denoted
by $\Gamma^{(\gamma)}_{\mu }$, is given by \cite{3}

\begin{eqnarray}
\Gamma^{(\gamma)}_{\mu}(q^2)=& i \gamma_{\mu}
\sum_{i=0,3,8} \lambda_{i} {\cal G}^{(\gamma)}_{i}(q^2)& \nonumber \\
 =&i \frac{e}{2}\gamma _{\mu} [ (\lambda _3 +
\frac{\lambda _8}{\sqrt{3}})
 + \sum _{i,j=0,3,8} & \lambda _i (J_{3j}(q^2) +
J_{8j}(q^2)) {\cal M}_{ji}(q^2)]
\label{e2}
\end{eqnarray}

The quantities $J_{3j}$ and $J_{8j}$ are the fundamental
quark bubbles which couple the bare electromagnetic
vertex to the vector particles $\rho $, $\omega $ and
$\phi $. These particles are generated through the coupling to
the strong quartic $q \bar{q}$ interaction, represented
by the $3 \times 3$ matrices ${\cal M}_{ji}$,
where i and j are flavor indices. The electromagnetic
Ward identity for this vertex is fulfilled. In the
present case, with the SU(2) flavor symmetry preserved
in the up and down sector, the $\rho$ - meson decouples
from the $\omega$ - $\phi$ excitations and ${\cal M}_{ji}$
reduces to a number in the case of the $\rho$ - meson and
to a $2 \times 2$ submatrix which describes $\omega$ -
$\phi$ mixing. The off-shell ${\rho}  q \bar{q}$
coupling ${\cal G}^{(\rho)}_{3}(q^2) $ can be extracted
from the $q \bar{q}$ scattering matrix ${\cal M}_{33}$
by separating the pole contribution and is given by

\begin{equation}
\Gamma^{(\rho)}_{\mu}(q^2)=i \gamma_{\mu} \lambda_{3}
{\cal G}^{(\rho)}_{3}(q^2) =
i \gamma_{\mu} \lambda_{3}
\sqrt{\frac{16}{9}G_2\frac{q^2-m_{\rho}^2}
{1-\frac{16}{9}G_2J_V(q^2)}}
\label{e3a}            
\end{equation}

\noindent
where $J_V $ is the quark bubble which appears in the
Bethe Salpeter solution to one-loop order in the
vector-isovector channel and $m_{\rho} $ the $\rho $ -
meson mass obtained at the corresponding pole. The
coupling is momentum dependent and contains all the
information about the $q \bar{q}$ structure of the
$\rho $ meson. In a similar way the ${\sigma} q \bar{q}$
couplings ${\cal G}_0^{(\sigma)}$ and ${\cal G}_8^{(\sigma)}$
are extracted from the corresponding $q \bar{q}$
scattering matrix; for details see e.g. \cite{2a}.
Following Feynman rules we obtain
for the $\sigma VV'$ amplitude, with $V,V'=\gamma$ or $\rho$

\begin{equation}
A_{\sigma VV'}=\epsilon_{3b}^\mu \epsilon_{2a}^\nu
\sum _{n=0,8} \sum _{i=u,d,s} M_i {\cal G}_n^{(\sigma)}
(q_1^2)X^{VV'}_{i,n}
(q_2^2,q_3^2) T_{\mu \nu}^{i}(q_2,q_3)
\label{e3}            
\end{equation}

\noindent
where $q_3=q_1+q_2$ and

\begin{equation}
T_i^{\mu \nu}=6i\int _{\Lambda} \frac{d^4k}{(2\pi)^4}
\frac{Tr_D\gamma^\mu(\, \kslash+M_i)
\gamma^\nu (\, \kslash-\qslash_2+M_i) {\bf 1}
(\, \kslash-\qslash_3+M_i)}
{[k^2-M_i^2][(k-q_2)^2-M_i^2][(k-q_3)^2-M_i^2]}
\label{e4}            
\end{equation}
\vspace{0.5cm}

The factors $X^{VV'}_{i,n}$ are the result of the flavor
trace in the triangle and involve the momentum dependent
couplings of the photon or $\rho $ to the quarks.
The factor 6 results from exchange and from taking the
trace over color. We have explicitly for $V$ and $V'$ being photons
\vspace{0.5cm}

\begin{equation}
X^{VV'}_{i,0}=
\sqrt{\frac{2}{3}}\sum _{n,m=0,3,8}{\cal G}_n^{(\gamma)}(q_2^2)
{\cal G}_m^{(\gamma)}(q_3^2)c_{nm}^i
\label{e4n}            
\end{equation}
with $c_{nm}^i=c_{mn}^i$ where
$c_{00}^u = \sqrt{2}c_{08}^u = 2c_{88}^u =
\frac{2}{3}c_{33}^u = \sqrt{2}c_{38}^u = c_{03}^u =\frac{2}{3},$
all $c_{nm}^d = c_{nm}^u$ except $c_{n3}^d = -c_{n3}^u$ $(n \neq 3),$ and
$c_{00}^s = -\frac{1}{\sqrt{2}}c_{08}^s = \frac{1}{2}c_{88}^s
 = \frac{2}{3}.$
The coefficients $c_{n3}^s$ equal zero. One has furthermore
$X^{VV'}_{i,0}=\frac{1}{\sqrt 2}X^{VV'}_{i,8};$ $(i=u,d)$ and
$X^{VV'}_{s,0}=-\frac{1}{2}X^{VV'}_{s,8}.$
\vspace{0.5cm}
In the case of one or both $V$ and $V'$ being a $\rho$ - meson one has to
substitute correspondingly ${\cal G}_n^{(\gamma)}$ by the $\rho$-$q \bar{q}$
coupling ${\cal G}_3^{(\rho)}$ and there is obviously no
triangle with strange quarks involved in (\ref{e3}).
\vspace{0.5cm}

The expression (\ref{e3}) for the vertex function
$A_{\sigma VV'}$ can be written in a more compact form

\begin{equation}
A_{\sigma VV'}=\epsilon_{3b}^\mu \epsilon_{2a}^\nu
T_{\mu \nu}(q_2,q_3)
\label{e4b}            
\end{equation}

In order to evaluate $T_{\mu \nu}$ we decompose in gauge
invariant tensors \cite{10} as

\begin{equation}
T_{\mu \nu} = F_{TT}G_{\mu \nu} + F_{LL}L_{\mu \nu}
\label{e5}            
\end{equation}

\noindent
with orthogonal projection operators
\begin{equation}
G_{\mu \nu}=g_{\mu \nu}+(q_3^2q_{2\mu}q_{2\nu}
+q_2^2q_{3\mu}q_{3\nu}-q_2q_3
(q_{2\mu}q_{3\nu}+q_{3\mu}q_{2\nu}))/Y
\label{e6}            
\end{equation}

\begin{equation}
L_{\mu \nu}=q_2q_3\frac{1}{Y} Q_{3\mu} Q_{2\nu}
\label{e7}            
\end{equation}

\noindent
where $Q_{3\mu}=q_{3\mu}-(q_3^2/q_3q_2) q_{2\mu}\,$,
$Q_{2\nu}=q_{2\nu}-(q_2^2/q_3q_2) q_{3\nu}\,$ and
$Y = (q_2q_3)^2-q_2^2q_3^2$.

\vspace{0.5cm}
{}From eqs. (\ref{e3}), (\ref{e4b}) and (\ref{e5}) it is easy to see
that

\begin{equation}
F_{TT(LL)}(q_1^2,q_2^2,q_3^2)= \sum _{n=0,8} \sum _{i=u,d,s}
M_i {\cal G}_n^{(\sigma)}(q_1^2)X^{VV'}_{i,n}(q_2^2,q_3^2)
F_{TT(LL)}^{i}(q_1^2,q_2^2,q_3^2).
\label{e7b}            
\end{equation}

In order to evaluate the transverse $F_{TT}$ and
longitudinal $F_{LL}$ form factors, we need first
the triangle contributions of each flavor separately,
which are obtained with the contractions

$$F_{TT}^{i}=\frac{1}{2}G_{\mu \nu }T^{\mu \nu}_{i}$$

\noindent
and

$$F_{LL}^{i}=\frac{(q_3q_2)^2}{q_3^2q_2^2}L_{\mu \nu} T^{\mu \nu}_{i}.$$

\vspace{0.5cm}

The amplitude $T_{\mu \nu}$ from (\ref{e4b}) is calculated
with the following prescription (which is in a sense a definition
of the cutoff): first, the reduction methods of ref.\cite{11}
are used; the integral is treated at this stage as without cutoff,
allowing in this
way translation of the integration variable. As a result,
$T_{\mu \nu}$ is expressed as a linear combination
of the basic integrals of the form (\ref{e9}) below
and in this way the $\mu,\nu$ dependence of the
integration variable removed. Second, a sharp Euclidean
cutoff is introduced for the scalar integrals $I_{2}$ and
$I_{3}$ (\ref{e9}). All the cutoff dependence of $T_{\mu \nu}$
thus resides in these basic integrals. The result could be in
principle gauge dependent, due to the introduction of a sharp
Euclidean cutoff. A necessary condition for
gauge independence of $T_{\mu \nu}$ is
that the Ward identities $q_{3}^{\mu} T_{\mu \nu}= q_{2}^{\nu}
T_{\mu \nu}=0$ are fulfilled. We can show, that these Ward
identities are indeed exactly fulfilled, regardless
of the cutoff $\Lambda$, with no need to
specify what $I_{2}$ and $I_{3}$ are (the coefficients in
front of $I_{2}$ and $I_{3}$ in the expressions
$q_{3}^{\mu} T_{\mu \nu}$ and $q_{2}^{\nu} T_{\mu \nu}$
vanish exactly). In this
way we check that the above explained recipe for handling
the initial expression (\ref{e4b}) is meaningful.

The results for the form factors can be cast in the form

\begin{equation}
F_{TT}^{i}=\frac{1}{2}[{\cal A}_{i}-\frac{q_2q_3}{Y}{\cal H}_{i}]
\label{e8}            
\end{equation}

\noindent
and

$$F_{LL}^{i}=\frac{q_2q_3}{Y}{\cal H}_{i}$$

\noindent
with

$${\cal A}_{i} = g_{\mu \nu}T^{\mu \nu}_{i} = $$
$$4(I_3^{i}\times (4q_2q_3+4M_{i}^2-q_3^2-q_2^2)+
I_2^{i}(q_2^2)+ I_2^{i}(q_3^2)-
2I_2^{i}((q_2-q_3)^2)))$$

\noindent
and

$${\cal H}_{i} = q_{2\mu}q_{3\nu}T^{\mu \nu}_{i} = $$
$$4((q_2q_3-q_2^2)(q_2q_3I_3^{i} + I_2^{i}(q_2^2
))+(q_2q_3-q_3^2)(q_2q_3I_3^{i} + I_2^{i}(q_3^2))+(q_2-q_3)^2
I_2^{i}((q_2-q_3)^2)))$$

\noindent
with the standard integrals

\begin{equation}
I_2^{i}(q^2)=i\int _{\Lambda} \frac{d^4k}{(2\pi)^4}
\frac{1}{[k^2-M_i^2] [(k+q)^2-M_i^2]}
\label{e9}            
\end{equation}

\noindent and
$$I_3^{i} = I_3^{i}(q_2^2,q_3^2,q_2q_3)=$$
$$i\int _{\Lambda} \frac{d^4k}{(2\pi)^4} \frac{1}
{[k^2-M_i^2][(k-q_2)^2-M_i^2][(k-q_3)^2-M_i^2]}.$$
\vspace{0.5cm}

\section{RESULTS AND DISCUSSION}
\vspace{0.5cm}
With the choice of parameters of ref.\cite{2b}
we obtain the masses $m_{\sigma}=775$ MeV and
$m_{\rho}=775$ MeV. Let's start to discuss the
results for the $\sigma \gamma \gamma$ vertex,
which may serve as a guide to extract information
about the existence of a broad resonant state in
the reaction $\gamma \gamma \rightarrow \ 2\pi$.
The interesting kinematical region is with two
on-shell photons and a time-like scalar near
its on-shell value. The form factor $F_{TT}$
for this case is displayed in fig.2.
The longitudinal form factor $F_{LL}$ is for
on-shell photons of course irrelevant,
because they cannot be longitudinal.
An interesting feature can be seen for the
transversal form factor $F_{TT}$: it is very
small for an on-shell $\sigma$-meson. It can be
proved analytically from eq. (\ref{e8}) and the
definitions for ${\cal A}_{i}$ and ${\cal H}_{i}$
that $F_{TT}^i=2(4M_i^2-q_1^2)I_3^i$ for real photons.
In the exact chiral $U(3)_{L} \times U(3)_{R}$
symmetry limit the constituent masses of all
the u, d and s quarks are equal and one
obtains the relation $m_{\sigma}=2 M_{u}$.
In this limit $F_{TT}$ for $\sigma \gamma \gamma$
vanishes exactly  for on-shell photons
and $\sigma$. In our case the chiral
$U(3)_{L} \times U(3)_{R}$ symmetry is
explicitly broken by $m_{curr} \ne 0$
and $K \ne 0$, so that the form factor
$F_{TT}$ is not exactly zero but close to it.
This means,
that the contribution of the $\sigma$ to the
process $\gamma \gamma \rightarrow \pi \pi$
is very small close to the $\sigma$ mass. We will discuss in the sequel
that
the zero of the relevant on-shell form factor
for $\sigma \gamma \gamma$ in the chiral
limit is the underlying physics
which forbids the observation of a scalar
resonance in processes such as
$\gamma \gamma \to \pi \pi$ and
$\rho \to \gamma (\pi \pi)_{S-wave}$. It
will explain how a scalar excitation can
exist, but not contribute to
certain processes. To obtain the complete NJL prediction to the
$\gamma \gamma \to \pi \pi$ reaction via scalar exchange, one needs to
couple the $\sigma \gamma \gamma$ vertex to the two pions in the exit
channel, taking in account the decay
width of the scalar into pions. This
corresponds to the 'resonant' contribution.
For the model parameters used
the coupling of the NJL scalar to two pions
yields a broad resonance width
$\Gamma_{\sigma \pi \pi}$ = 435MeV at the
$\sigma$ mass.
A simple estimate of the cross section for
this process can be obtained
assuming a Breit-Wigner type resonance, with
the width as a function of the square momentum of
the $\sigma$ predicted in NJL.
The contribution of the $\sigma$ meson should
not be large, since it would otherwise
enlarge too much the experimentally small cross--section
in the $\pi^{0} \pi^{0}$ channel \cite{exppi0pi0}.
We found that within the NJL and with the Breit-Wigner
form for the  $\sigma$ propagator the resonant
contribution is very small near the scalar mass
($\approx 0.3$ nb at $W_{\gamma \gamma}=0.775$ GeV),
but is increasing with decreasing energy
($\approx 10$ nb at $0.5$ GeV) until it becomes too large
($\approx 50$ nb at $0.32$ GeV).
Nevertheless, near the sigma mass, the resonant
scalar contribution is shadowed by chiral
symmetry, so that the existence of the $\sigma$
is not in contradiction with the data in this
kinematical region.

A comment is in order concerning a complete calculation of
the $\gamma \gamma \rightarrow \pi^0 \pi^0$ reaction within
the NJL model: processes other than the scalar exchange
considered presently are involved to the same order,
like a quark box diagram with two photons and two pions attached
to the vertices, and the exchange of vector mesons in the $u$ and $t$ -
channels. These processes can be denoted as interference or
non-resonant terms. However, for the purposes of showing that the
predicted broad NJL scalar is compatible with the small cross sections
observed below the nominal $\sigma$ mass in the reaction
$\gamma \gamma \rightarrow \pi^0 \pi^0$, it is enough to consider the
"resonant" term for the following reasons. We
calculate the vector meson exchange for $cos(\theta) = 0$
and find it to be negligible in the kinematical region
considered. We expect the box diagram (which we have not
calculated) to lower the cross section, since the $\sigma$
exchange graph and the box diagram cancel in the chiral
limit and for photon
frequencies $\omega_i \rightarrow 0$, see \cite{12}.
This fact is especially welcome near
threshold, where the resonant contribution predicted
by the NJL is too large. A complete calculation is
rather involved and relegated to a later work.
In our opinion it will not alter the conclusion that
chiral symmetry is at work producing a small
$\sigma \gamma \gamma$ amplitude, which
vanishes in the chiral limit (due to $m_{\sigma}=2M_u$)
and is responsible for the
small resonant scalar contribution
observed around the nominal $\sigma$ mass.

The screening of the $\sigma$ contribution is
present also in the
$\rho \to \gamma (\pi \pi)_{(S-wave)}$ decay. The
intermediate $\sigma$ must be somewhat off-shell
because of phase space. From fig. 3 it
is in fact possible to see that the $F_{TT}$
of the $\sigma \rho \gamma$ vertex
is small whenever the scalar and
the photon are nearly on-shell.
In the case of one on-shell photon
the transverse form factors reduce to

$$F_{TT}^i=2((4M_i^2-q_1^2+q_2^2)I_3^i
-\frac{2q_2^2}{q_2^2-q_1^2}
(I_2^i(q_2^2)-I_2^i(q_1^2)))$$

\noindent
as can be obtained again from (\ref{e8}).
One observes in fig. 3 that $F_{TT}$ is
almost insensitive to the variations in
$q_2^2$ of the $\rho$-meson at the
on-shell values of the scalar and photon.
Since $F_{TT}$ is close to 0 for on-shell
$\sigma$ and $q_{2}^{2}=0$ (similar situation as
the one discussed above for $\sigma \gamma
\gamma$), it remains close to 0 also for a
nearly on-shell $\rho$-meson. In the same figure we
notice a steep increase of the form factors as
functions of increasing absolute values of
the squared space-like momenta of the
scalar. The exchange of a
{\it space-like} $\sigma$ in the diagram with
two $\sigma \rho \gamma$ vertices can give
a sizeable contribution to the process $\gamma
\gamma \to \rho^{0} \rho^{0}$, see fig.3.

Significant changes are also seen in the
$\sigma \rho \rho$ form factors as shown in figs. 4a
and 4b. These form factors are also relevant for the
$\rho \rho $ interaction via scalar exchange. Since
the form factors drop rather rapidly at the on-shell
values of the $\rho$ -mesons near $q_1^2 \simeq 0$ as functions of
the $q_1^2$ of the $\sigma$ meson
, the Fourier transform of $F_{TT}^2 /
(q^2-m_{\sigma}^2)$ will be shallower and of longer
range than the corresponding Yukawa potential. However
such effective potentials are dependent on the off-shell
mass of the $\rho$-mesons (see figs.4); for lower $q^2$
of the $\rho$ the potential becomes deeper and of shorter
range. This might explain, why the enhancement remains
large even below the nominal $\rho^{0}\rho^{0}$ threshold.
Namely, "lighter" $\rho$-mesons (off-shell $\rho$-mesons)
would feel a deeper potential and "resonate" at the
lower energy. This effect might make the resonance
"follow" into the region below $1.5$ GeV, rendering it
broader and noticeable below $1.5$ GeV.

\vspace{0.5cm}

\section{CONCLUSIONS}

\vspace{0.5cm}
To summarize, we have shown that the $\simeq 400$MeV
wide scalar resonance predicted by the NJL model is
a good candidate to describe several radiative and
strong interaction phenomena. In particular we have discussed the
remarkable fact that the presence of the $\sigma$ excitation in the
reactons $\gamma \gamma \to \pi \pi$ and
$\rho \to \gamma (\pi \pi)_{(S-wave)}$ is hidden due to chiral
symmetry. This results are a beautiful illustaration of chiral
symmetry at work at the scale of the $\sigma$ mass.
In the case of the reaction $\gamma \gamma \to \rho
\rho \to 4 \pi$, where
a large enhancement of the $\rho^0 \rho^0$ intermediate
channel as compared to $\rho^+ \rho^-$ is observed, we have
proposed a picture in which the reaction
proceeds via a scalar exchange. We have calculated
the vertices of relevance for
this process, $\sigma \rho \gamma$ and
$\sigma \rho \rho$ and discussed the
reasons for the enhancement. The vertices
involving the $\sigma$ meson
calculated here can be used in the evaluation
of cross sections. Work in
this direction is in progress.
\vspace{0.5cm}

{\bf Acknowledgement}
We thank Matja\v z Polj\v sak for many useful
discussions.

\vspace{0.5cm}

{\bf FIGURE CAPTIONS}
\vspace{0.5cm}

\noindent fig.1 Diagram to calculate the vertex $\sigma VV'$;
V,V'=photon or vector meson (wavy lines) with momenta $q_2$
and $q_3$. The $\sigma $ - mode has momentum $q_1$. The large circles with
crosses and the full circle are to remind that the vertices are momentum
dependent.
\vspace{0.5cm}

\noindent fig.2 Transverse $F_{TT}$ form factor for the vertex
$\sigma \gamma \gamma$
with both photons on-shell, as function of $q_1^2$ of
the scalar.
\vspace{0.5cm}

\noindent fig.3 $F_{TT}$ for $\sigma \rho \gamma$ with
on shell photon.
The abscissa is the off-shell mass squared of the scalar.
Curves: a) with $\rho$-meson on shell;
b) $ q_2^2  = .5$ GeV$^2$; c) $ q_2^2  = .3$ GeV$^2$.
\vspace{0.5cm}

\noindent fig.4a $F_{TT}$ for $\rho \rho \sigma$.
The abscissa is the off-shell mass squared of the
scalar. Curves: a) off-shell masses
of the  $\rho $ - mesons $ q_2^2 = q_3^2 = .6$ GeV$^2$;
b) $ q_2^2 = .5$ GeV$^2$, $q_3^2 = .6$ GeV$^2$;
c) $ q_2^2 = q_3^2 = .5$ GeV$^2$
\vspace{0.5cm}

\noindent fig.4b Same as in 4a but for $F_{LL}$.
\vspace{0.5cm}

\end{document}